\documentclass[prx, english, twocolumn, longbibliography]{revtex4-1}
\usepackage{bm, amsmath, amsfonts, amssymb, braket,mathtools,upgreek,amsthm}

\DeclareMathOperator{\adj}{adj}
\usepackage{multirow}
\usepackage{float}
\usepackage{color}
\usepackage{comment}
\usepackage{subfigure}
\usepackage{lipsum}

\allowdisplaybreaks

\begin{document}

\newcommand{\ii}{\text{i}}
\newcommand{\U}{U}
\newcommand{\V}{V}
\newcommand{\BZ}{\left[ 0, 2\pi \right]}
\newcommand{\CZ}{\left[ 0, 1 \right]}

\title{Exact formulas of the end-to-end Green's functions in non-Hermitian systems}

\author{Haoshu Li}
    \email{lihaoshu@mail.ustc.edu.cn}
    \affiliation{Department of Modern Physics, University of Science and Technology of China, Hefei 230026, China}

\author{Shaolong Wan}
    \email{slwan@ustc.edu.cn}

    \affiliation{Department of Modern Physics, University of Science and Technology of China, Hefei 230026, China}

\begin{abstract}
    Green's function in non-Hermitian systems has recently been revealed to be capable of directional amplification in some cases. The exact formulas for end-to-end Green's functions are significantly important for studies of both non-Hermitian systems and their applications. In this work, based on the Widom's formula, we derive exact formulas for the end-to-end Green's functions of single-band systems which depend on the roots of a simple algebraic equation. These exact formulas allow direct and accurate comparisons between theoretical results and experimentally measured quantities. In addition, we verify the prior established integral formula in the bulk region to agree with the result in our framework. We also find that the speed at which the Green's functions in the bulk region approach the prior established integral formula is not slower than an exponential decay as the system size increases. The correspondence between the signal amplification and the non-Hermitian skin effect is confirmed. 
\end{abstract}

\maketitle

\section{Introduction}
Hermitian Hamiltonians are traditionally the focus of quantum physics and condensed matter research. There is a surge in interest in systems with non-Hermitian Hamiltonians in recent years. Open quantum systems \cite{malzard2015, open1, open2, open3, open4, open5} and optical systems subjected to gain and loss \cite{gain1, gain2, gain3, gain4, gain5, gain6, gain7, gain8, gain9, gain10, gain11, gain12} could be the source of these systems. The non-Hermitian skin effect (NHSE) \cite{PhysRevLett.116.133903, yao2018, yao20182, PhysRevB.97.121401, PhysRevLett.121.026808, PhysRevX.8.031079, Thomale2019, londhi2019, song2019, PhysRevResearch.1.023013, origin2020} has been studied as a result of recent advancements in non-Hermitian physics. In contrast to Hermitian instances, the majority of eigenfunctions of a non-Hermitian Hamiltonian are localized at the lattice boundary, implying that the conventional bulk-boundary correspondence is broken in non-Hermitian systems. Non-Hermitian systems can also display the high-order skin effect \cite{Edvardsson2019, Ezawa2019, kawabaras, zhang2019, lee2019, denner2020, okugawa, PhysRevB.102.205118, PhysRevB.103.045420}. The notion of the generalized Brillouin zone (GBZ) \cite{yao2018, yao20182, yokomizo2019, PhysRevLett.125.226402, PhysRevB.100.035102, PhysRevLett.124.066602, PhysRevB.101.195147, PhysRevLett.123.246801, PhysRevB.102.085151, PhysRevLett.125.186802, RevModPhys.93.015005, doi:10.1080/00018732.2021.1876991} is developed in the research of NHSE, and this concept can be utilized to compute energy spectra under open boundary conditions (OBCs). The GBZ is the unit circle in the complex plane in the Hermitian case, and a collection of closed loops in the complex plane in the non-Hermitian situation. Recently, there is a growing body of literature that recognizes the importance of the Greens' functions in non-Hermitian systems \cite{PhysRevLett.124.056802, PhysRevLett.126.216407, PhysRevB.99.125155, PhysRevB.103.195157, PhysRevLett.80.2897}. It is discovered that the Green's functions contain useful information about the bulk-boundary correspondence \cite{PhysRevLett.124.056802, PhysRevLett.126.216407}, the NHSE \cite{ PhysRevLett.126.216407, PhysRevB.103.195157}, and the quantized response associated with the spectral winding topology \cite{Li2021-1, Li2021-2}. Furthermore, the Green's function was shown to play a crucial role in the topological field theory of non-Hermitian systems \cite{PhysRevLett.126.216405}. In the study of Green's functions, GBZ technique has been used \cite{PhysRevB.103.L241408}. Because they can be employed to achieve directional amplification \cite{Wanjura2020, PhysRevX.5.021025, Ranzani_2015, PhysRevLett.122.143901, PhysRevX.3.031001, PhysRevLett.112.167701, PhysRevX.5.041020, Jalas2013, Feng729, PhysRevApplied.10.047001, PhysRevX.7.031001}, the end-to-end Green's functions are concentrated. For a signal $\mathbb{\epsilon}$ with a frequency $\omega$, $\mathbb{\epsilon}(t) = \mathbb{\epsilon}(\omega) \exp(-i \omega t)$ where $\mathbb{\epsilon} = (\epsilon_1,\ldots,\epsilon_L)^T$, the output signal is $\psi(t) = \psi(\omega) \exp(-i \omega t)$ with the amplitude in the frequency domain given by $\psi(\omega) = G(\omega) \mathbb{\epsilon}(\omega)$. In particular, the directional amplification of a signal input at one end of a one-dimensional (1D) chain and measured at the other end is described by the end-to-end Green's functions $G_{1,L}(\omega)$ and $G_{L,1}(\omega)$ \cite{Wanjura2020, PhysRevB.103.L241408, Li2021-1, Li2021-2}. Up to now, the exact formulas for the end-to-end Green's functions have still be not obtained. We believe that the exact formulas for the end-to-end Green's functions are significantly important for both studies of non-Hermitian systems and their applications.

The main aim of this research is to look at the analytic form of the 1D chain's Green's function under OBC, with a focus on the entry of the Green's function which can be used to achieve amplification \cite{PhysRevB.103.L241408}, i.e., $G_{1,L}(\omega)$ and $G_{L,1}(\omega)$, where $L$ is the length of the 1D chain. In order to obtain exact $G_{1,L}(\omega)$ and $G_{L,1}(\omega)$ formulas, we use the Widom's formula, and we believe that these exact formulas make it possible to directly and accurately compare theoretical results with quantitative experimental measurable quantities, which will be useful in future experiments. Furthermore, by using the Widom's formula and a generalized concept of circulant matrices, we verify the prior established integral formula in the bulk region of the 1D chain to agree with the result in our framework. We also find that the speed at which the Green's functions in the bulk region approach the prior established integral formula is not slower than an exponential decay as the system size $L$ increases, i.e., the Green's functions in the bulk region are displaced from the prior established integral formula by $O(e^{-b L})$.

The paper is organized as follows: In Sec.~\ref{sec: HN}, we derive the end-to-end Green's functions in the Hatano-Nelson (HN) model \cite{hn} which depend on the roots of a simple algebraic equation and show that all zeros of the characteristic function contribute to the end-to-end Green's functions of systems with finite size. In addition, for a large enough system size, the exact asymptotic formula is obtained in this section. In Sec.~\ref{sec: arb_range}, we analyze the single band model with arbitrary hopping range, develop the analytic formula of the end-to-end Green's functions, and calculate the asymptotic formula of the Green's function for a large enough system size. In Sec.~\ref{sec: main_green_bulk}, we verify the GBZ-based integral formula in the bulk region to agree with the result in our framework. In Sec.~\ref{sec: NHSE}, we discuss the relation between the signal amplification and the NHSE. In Sec.~\ref{sec: con}, we present our conclusions. In Appendix \ref{sec: range_two}, we offer the formula of the $(1, L)$ entry of the Green's function of the tight-binding model whose hopping range is $M=2$. In Appendix \ref{sec: another}, we derive the asymptotic formula of the $(L, 1)$ entry of the Green's function. In Appendix \ref{sec: diag}, we give the diagonal elements of the Green's function at the end of the chain. In Appendix \ref{sec: proof_bulk}, we prove the GBZ-based integral formula of Green's functions is accurate in the bulk region using an estimation building in Appendix \ref{sec: estimate}.

\section{HN model} \label{sec: HN}
We investigate the HN model \cite{hn}
\begin{align}
    H = \sum_{n=1}^{L-1} (t_{-1} c^{\dagger}_{n+1} c_n + t_1 c^{\dagger}_n c_{n+1}).
\end{align}
Taking the OBC, the single-particle Hamiltonian is given by
\begin{align}
    H = \begin{pmatrix}
        0 & t_1 & 0 & \cdots & \cdots & 0 \\
        t_{-1} & 0 & t_1 & 0  & \cdots & 0 \\
        0 & t_{-1} & 0 & t_1 & \cdots & 0 \\
        \vdots & \ddots & \ddots & \ddots & \ddots & \vdots \\
        0 & \cdots & 0 & t_{-1} & 0 & t_1 \\
        0 & \cdots & \cdots & 0 & t_{-1} & 0
    \end{pmatrix}_{L \times L} .
\end{align} 

The frequency-domain Green's function of this system is $G(\omega) = (\omega-H)^{-1}$, where $\omega - H$ is given by
\begin{align}
    \omega - H =\begin{pmatrix}
        \omega & -t_1 & 0 & \cdots & \cdots & 0 \\
        -t_{-1} & \omega & -t_1 & 0  & \cdots & 0 \\
        0 & -t_{-1} & \omega & -t_1 & \cdots & 0 \\
        \vdots & \ddots & \ddots & \ddots & \ddots & \vdots \\
        0 & \cdots & 0 & -t_{-1} & \omega & -t_1 \\
        0 & \cdots & \cdots & 0 & -t_{-1} & \omega
    \end{pmatrix}_{L \times L} .
\end{align}
Using Cramer's rule for finding the inverse matrix, the $(1, L)$ entry of the Green's function is given by
\begin{align} \label{eq: green}
    G(\omega)_{1, L} & = (\omega-H)^{-1}_{1, L} \notag \\
    & = \frac{\adj(\omega-H)_{1,L}}{\det(\omega-H)} \notag \\
    & = \frac{(-1)^{L-1} \begin{vmatrix}
        -t_1 & 0 & \cdots & 0 \\
        \omega & -t_1 & 0 & \cdots \\
        \vdots & \ddots & \ddots & \vdots \\
        0 & \cdots & \omega & -t_1
    \end{vmatrix}_{(L-1) \times (L-1)}}{\det (\omega-H)} \notag \\
    & = \frac{t_1^{L-1}}{\det(\omega-H)} ,
\end{align}
where $\adj(\omega-H)$ is the adjugate matrix of $\omega-H$.
As a result, to calculate $G(\omega)_{1, L}$, it is sufficient to calculate the determinant of $\omega-H$.

Denoting the following determinant by $\Delta_n$
\begin{align}
    \begin{vmatrix}
        \omega & -t_1 & 0 & \cdots & \cdots & 0 \\
        -t_{-1} & \omega & -t_1 & 0  & \cdots & 0 \\
        0 & -t_{-1} & \omega & -t_1 & \cdots & 0 \\
        \vdots & \ddots & \ddots & \ddots & \ddots & \vdots \\
        0 & \cdots & 0 & -t_{-1} & \omega & -t_1 \\
        0 & \cdots & \cdots & 0 & -t_{-1} & \omega
    \end{vmatrix}_{n \times n} =: \Delta_n.
\end{align}
Then 
\begin{align}
    \Delta_n = & \omega \Delta_{n-1} \notag  \\ 
    & + (-1) (-t_1)  \begin{vmatrix}
        -t_{-1} & -t_1 & 0 & \cdots & 0 \\
        0 & \omega & -t_1 & 0 & \cdots \\
        0 & -t_{-1} & \omega & -t_1 & \cdots \\
        \vdots & \ddots & \ddots & \ddots & \vdots \\
        0 & \cdots & 0 & -t_{-1} & \omega
    \end{vmatrix}_{(n-1) \times (n-1)} \notag \\
    = & \omega \Delta_{n-1} - t_1 t_{-1} \Delta_{n-2} .
\end{align}
With the initial conditions that $\Delta_1 = \omega$ and $\Delta_2 = \omega^2 -t_1 t_{-1}$, this second order difference equation can be easily solved. Its general solution is 
\begin{align} \label{eq: det}
    \Delta_n = a_1 z_1^{n-1} + a_2 z_2^{n-1},
\end{align}
where $z_1$ and $z_2$ are two roots of the characteristic polynomial $z^2 = \omega z - t_1 t_{-1}$, and 
\begin{align} \label{eq: a12}
    a_1 & = \frac{\omega}{2} - \frac{\omega^2 - 2 t_1 t_{-1}}{2 \sqrt{\omega^2 - 4 t_1 t_{-1}}}, \notag \\
    a_2 & = \frac{\omega}{2} + \frac{\omega^2 - 2 t_1 t_{-1}}{2 \sqrt{\omega^2 - 4 t_1 t_{-1}}} .
\end{align}

Inserting Eq.~(\ref{eq: det}) in Eq.~(\ref{eq: green}) we obtain
\begin{align} \label{eq: green_0}
    G(\omega)_{1, L} & = \frac{t_1^{L-1}}{\Delta_{L}} \notag \\
    & = \frac{1}{a_1 (\frac{z_1}{t_1})^{L-1} + a_2 (\frac{z_2}{t_1})^{L-1}} .
\end{align}
Note that 
\begin{align}
    \frac{z_1}{t_1} & = \frac{\omega - \sqrt{\omega^2 - 4 t_1 t_{-1}}}{2 t_1} = \beta_1 , \notag \\
    \frac{z_2}{t_1} & = \frac{\omega + \sqrt{\omega^2 - 4 t_1 t_{-1}}}{2 t_1} = \beta_2 ,
\end{align}
where $\beta_1$ and $\beta_2$ are two zeros of the function $t_1 \beta - \omega + t_{-1} \beta^{-1}$.

\begin{figure} 
    \subfigure[]{  
        \centering
        \includegraphics[width=0.4\columnwidth]{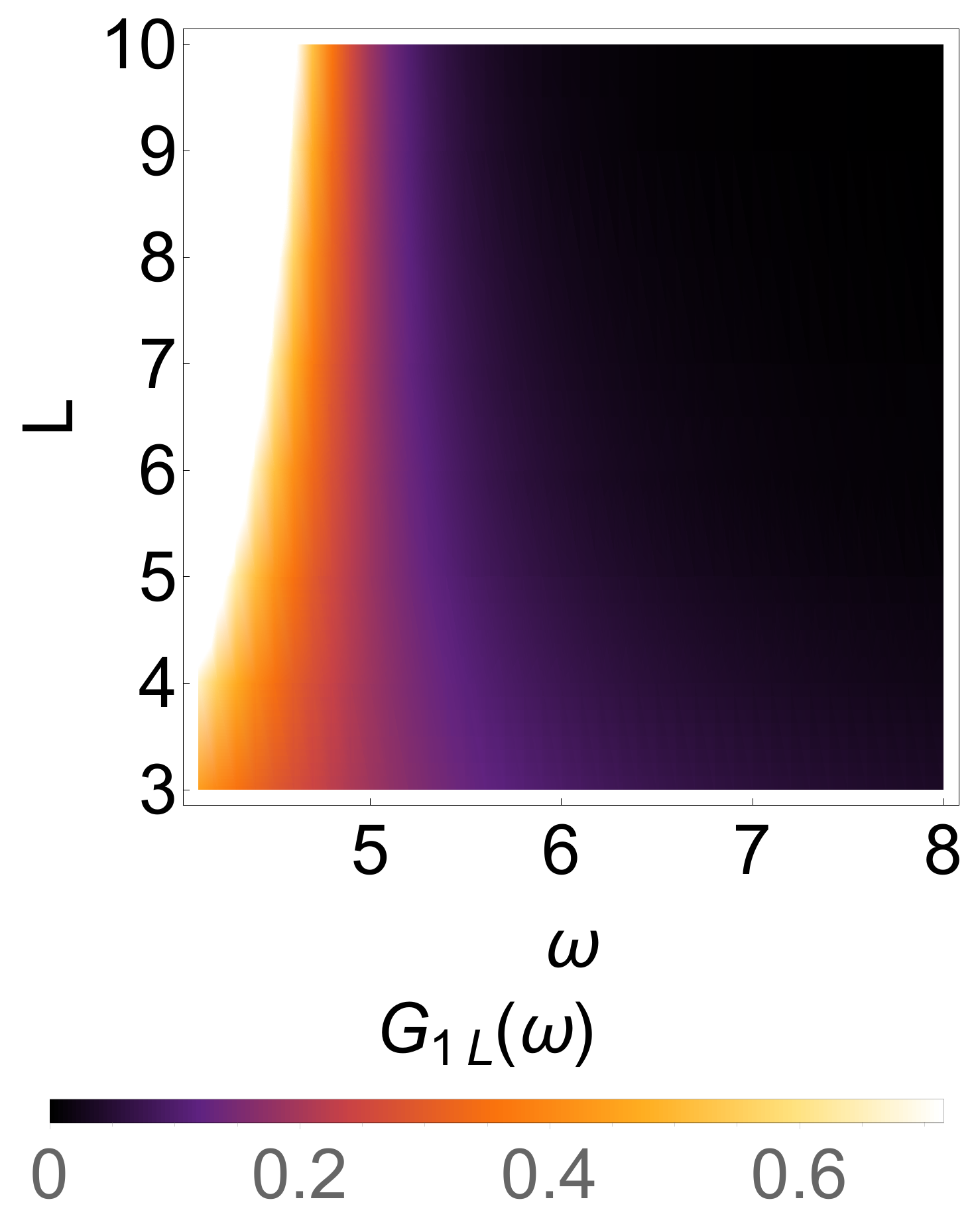}
    } 
    \subfigure[]{
        \centering
        \includegraphics[width=0.4\columnwidth]{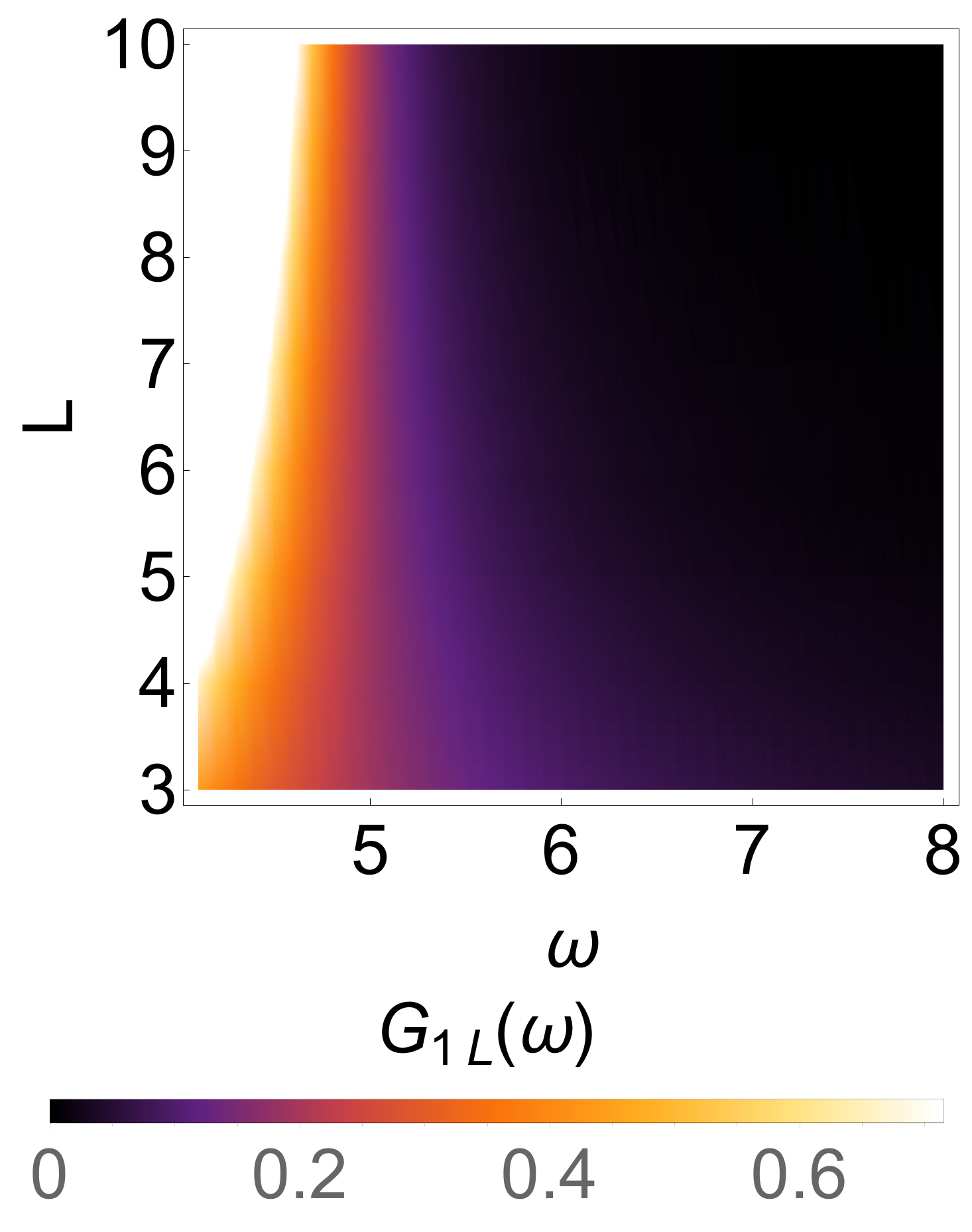}
    } 
    \caption[]{Let $t_1=4$ and $t_{-1}=1$. (a) $G(\omega)_{1, L}$ as a function of $\omega$ and $L$, as determined by Eq.~(\ref{eq: green_1}). (b) $G(\omega)_{1, L}$ as a function of $\omega$ and $L$, which is calculated numerically. }  \label{fig: density}
\end{figure}

Therefore, Eq.~(\ref{eq: green_0}) reduces to
\begin{align} \label{eq: green_1}
    G(\omega)_{1, L} = \frac{1}{a_1 \beta_1^{L-1} + a_2 \beta_2^{L-1}}.
\end{align}
Consider a case that $\omega > 2 \sqrt{t_1 t_{-1}} > 0$, then $|\beta_1| < \sqrt{\frac{t_{-1}}{t_1}} < |\beta_2|$, implying $\beta_1$ is on the inside of the GBZ and $\beta_2$ is on the outside of the GBZ. Hence all zeros of the function $t_1 \beta - \omega + t_{-1} \beta^{-1}$ contribute to the $(1,L)$ entry of the Green's function. Using Eq.~(\ref{eq: green_1}) and the numerical method, we plot $G(\omega)_{1, L}$ as a function of $\omega$ and $L$ in Fig.~\ref{fig: density}.

\begin{figure} 
    \centering
    \includegraphics[width = 0.8\columnwidth]{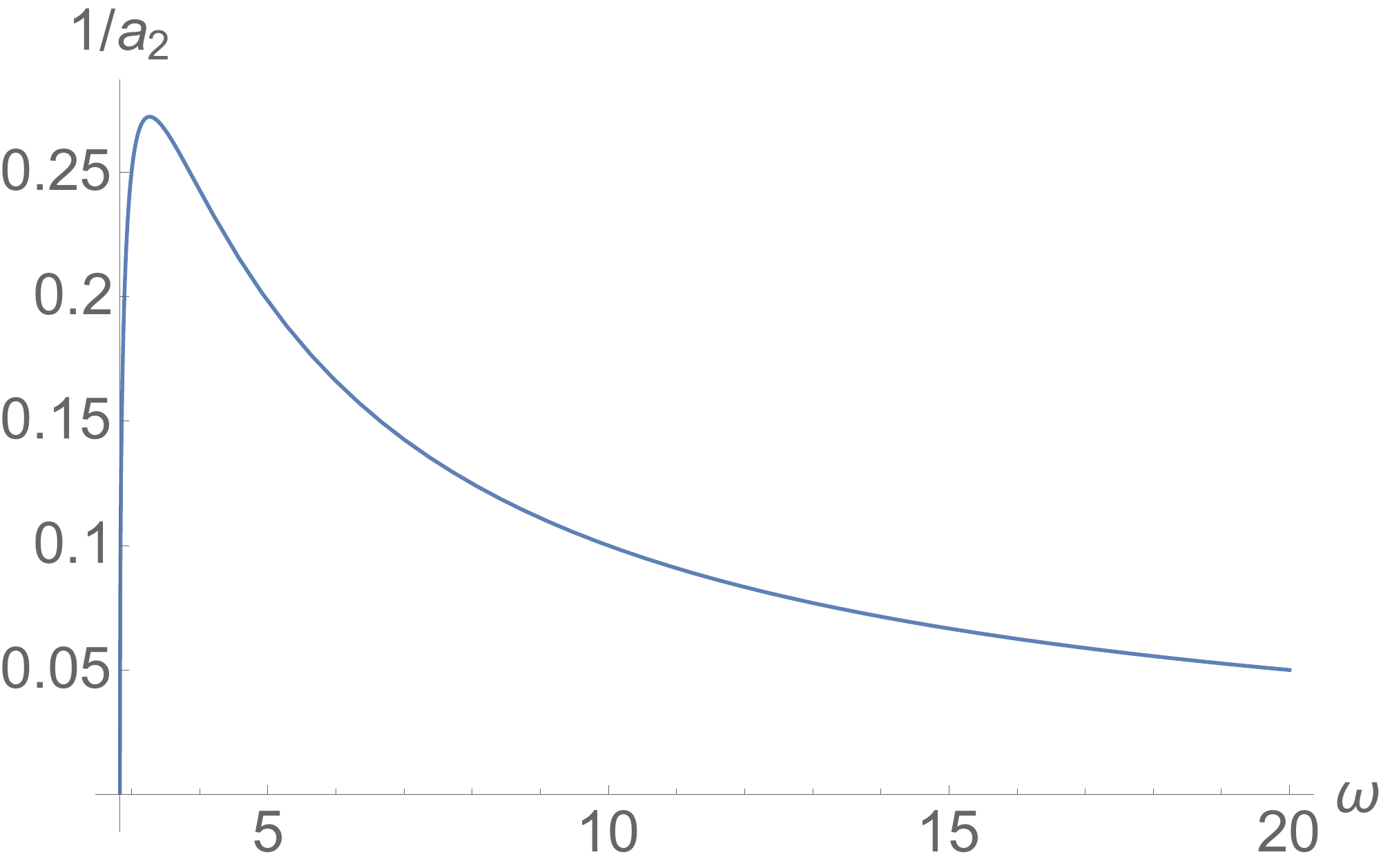}
    \caption{The coefficient $\frac{1}{a_2}$ as a function of the frequency  $\omega$. We take $t_{-1}=2$ and $t_1 =1$.} \label{fig: coeff}
\end{figure}

Now, we consider the asymptotic behavior of $G_{1, L}(\omega)$, since $|\beta_1| < |\beta_2|$, for a large system size $L$, we can get the exact asymptotic formula as
\begin{align} \label{eq: green_3}
    G_{1, L}(\omega) = \frac{1}{a_2} \beta_2^{-(L-1)},
\end{align}
where $a_2$ is given in Eq.~(\ref{eq: a12}).
Let $t_{-1}=2$ and $t_1 =1$, we plot the coefficient $\frac{1}{a_2}$ as a function of the frequency  $\omega$ in Fig.~\ref{fig: coeff}.

\section{Case of arbitrary hopping range} \label{sec: arb_range}
In this section, we look at a tight-binding model with a hopping range of $M$, which means that the hopping parameters are $t_{-M}, t_{-M+1},\ldots, t_{M-1}, t_{M}$.
The real-space Hamiltonian under OBC in this case is
\begin{align}
    H = \begin{pmatrix}
        t_0 & t_1 & \cdots & t_M & 0 & \cdots & 0 \\
        t_{-1} & t_0 & t_1 & \cdots & t_M & 0 & \cdots \\
        \vdots & \ddots & \ddots & \ddots & \ddots & \ddots & \vdots \\
        t_{-M} & t_{-M+1} & \cdots & \cdots & \cdots & \cdots & 0 \\
        \vdots & \ddots & \ddots & \ddots & \ddots & \ddots & \vdots \\
        0 & \cdots & 0 & t_{-M} & t_{-M+1} & \cdots & t_0
    \end{pmatrix}_{L \times L};
\end{align}
therefore
\begin{align}
    & \omega - H \notag \\
    = & \begin{pmatrix}
        \omega-t_0 & -t_1 & \cdots & -t_M & 0 & \cdots & 0 \\
        -t_{-1} & \omega-t_0 & -t_1 & \cdots & -t_M & 0 & \cdots \\
        \vdots & \ddots & \ddots & \ddots & \ddots & \ddots & \vdots \\
        -t_{-M} & -t_{-M+1} & \cdots & \cdots & \cdots & \cdots & 0 \\
        \vdots & \ddots & \ddots & \ddots & \ddots & \ddots & \vdots \\
        0 & \cdots & 0 & -t_{-M} & -t_{-M+1} & \cdots & \omega-t_0
    \end{pmatrix}.
\end{align}
The $(1,L)$ entry of the Green's function is
\begin{align} \label{eq: gen_green}
    & G_{1, L}(\omega) \notag \\
    = & (\omega-H)^{-1}_{1, L} \notag \\
    = & \frac{(-1)^{L-1} \begin{vmatrix} 
        -t_1 & \cdots & -t_M & \cdots & 0 \\
        \omega-t_0 & -t_1 & \cdots & \cdots & 0\\
        -t_{-1} & \omega-t_0 & \cdots  & \cdots  & 0 \\
        \vdots & \ddots & \ddots & \ddots & \vdots \\
        -t_{-M} & -t_{-M+1} & \cdots & \cdots & 0 \\
        \vdots & \ddots & \ddots & \ddots & \vdots \\
        0 & \cdots & -t_{-M} & \cdots & -t_1 
    \end{vmatrix}_{(L-1) \times (L-1)}}{\det(\omega-H)} .
\end{align}
Because the numerator and the denominator of RHS of Eq.~(\ref{eq: gen_green}) are both determinants of a Toeplitz matrix, a formula for the determinant of the Toeplitz matrix is required to calculate RHS of Eq.~(\ref{eq: gen_green}). The Widom's formula Eq.~(\ref{eq: det_toe}) \cite{bottcher2005spectral} is one such formula; it expresses the determinant of any Toeplitz matrix in terms of roots of the corresponding Laurent polynomial of the matrix and the size of the matrix. We'll just go over the results here.

Let $b$ be a Laurent polynomial,
\begin{align}
    b(z) = \sum_{j=-r}^{s} b_j z^j.
\end{align}
We give the formula of the determinant of the Toeplitz matrix
\begin{align}
    T_n(b) = \begin{pmatrix}
        b_0 & b_1 & \cdots & b_s & 0 & \cdots & 0 \\
        b_{-1} & b_0 & b_1 & \cdots & b_s & 0 & \cdots \\
        \vdots & \ddots & \ddots & \ddots & \ddots & \ddots & \vdots \\
        b_{-r} & b_{-r+1} & \cdots & \cdots & \cdots & \cdots & \cdots \\
        \vdots & \ddots & \ddots & \ddots & \ddots & \ddots & \vdots \\
        0 & \cdots & 0 & b_{-r} & \cdots & b_0 & b_1 \\
        0 & \cdots & 0 & 0 & b_{-r} & \cdots & b_0
    \end{pmatrix}_{n \times n}.
\end{align}
We can write
\begin{align}
    b(z) = b_s z^{-r} \prod_{j=1}^{r+s} (z-z_j),
\end{align}
where $z_1$, $z_2$, $\ldots$ , $z_{r+s}$ are the roots of the polynomial
\begin{align}
    z^r b(z) = b_{-r} + b_{-r+1} z + \cdots + b_s z^{r+s} .
\end{align}
If the zeros $z_1$, $\ldots$ , $z_{r+s}$ are pairwise distinct then, the determinant of $T_n(b)$ can be expressed by 
\begin{align} \label{eq: det_toe}
    \det \boldsymbol{(} T_n(b) \boldsymbol{)} = \sum_M c_M \omega_M^n,
\end{align}
where the sum is over all $\begin{pmatrix} r+s \\ s \end{pmatrix}$ subsets $M \subset \{1, 2, \ldots, r+s \}$ of cardinality $|M|=s$ and, with $\bar{M} := \{1, 2, \ldots, r+s \} \setminus  M$,
\begin{align} \label{eq: det_toe_detail}
    \omega_M & := (-1)^s b_s \prod_{j \in M} z_j , \notag \\
    c_M & := \prod_{j \in M} z_j^r \prod_{\substack{j \in M \\ k \in \bar{M}}} (z_j - z_k)^{-1} .
\end{align}

This formula tells us that all zeros of the function $t_M \beta^M + \cdots + (t_0-\omega) + \cdots + t_{-M} \beta^{-M}$ contribute to the $(1, L)$ entry of the Green function. For the particular case of the tight-binding model with $M=2$ hopping range, see Appendix \ref{sec: range_two}.

Now, we derive the asymptotic formula of the $(1, L)$-entry of the Green's function. The asymptotic formula is an accurate enough analytic formula in circumstances when the chain is long enough. The dominating terms of the numerator and the denominator of RHS of Eq.~(\ref{eq: gen_green}) are all that is required to obtain the asymptotic formula. Assume that $|\beta_1| \leqslant |\beta_2| \leqslant \cdots \leqslant |\beta_M | < |\beta_{M+1}| < |\beta_{M+2}| \leqslant \cdots \leqslant |\beta_{2M}|$, where $\beta_1, \ldots, \beta_{2M}$ are the zeros of the function $t_M \beta^M + \cdots + (t_0-\omega) + \cdots + t_{-M} \beta^{-M}$.

By Eq.~(\ref{eq: det_toe}), the dominant term of the numerator of Eq.~(\ref{eq: gen_green}) is 
\begin{align}
    c_A \omega_A^{L-1}, \notag
\end{align}
where
\begin{align}
    \omega_A & = (-1)^{M-1} (-t_M) \beta_{M+2}  \cdots \beta_{2M}, \notag \\
    c_A & = (\beta_{M+2} \cdots \beta_{2M})^{M+1} \prod_{\substack{j \in \{M+2,\ldots,2M\} \\ k \in \{1,\ldots,M+1\} }} (\beta_j - \beta_k)^{-1}.
\end{align}
Denote the numerator of Eq.~(\ref{eq: gen_green}) by A, 
\begin{align}
    A = (-1)^{L-1} c_A \omega_A^{L-1}.
\end{align}
By Eq.~(\ref{eq: det_toe}), the dominant term of the denominator of Eq.~(\ref{eq: gen_green}) is 
\begin{align} \label{eq: denom}
    c_B \omega_B^L, \notag
\end{align}
where
\begin{align}
    \omega_B & = (-1)^M (-t_M) \beta_{M+1} \cdots \beta_{2M}, \notag \\
    c_B & = (\beta_{M+1} \cdots \beta_{2M})^M \prod_{\substack{j \in \{M+1,\ldots,2M\} \\ k \in \{1,\ldots,M\} }} (\beta_j - \beta_k)^{-1}.
\end{align}
Denote the denominator of Eq.~(\ref{eq: gen_green}) by B,
\begin{align}
    B = c_B \omega_B^L.
\end{align}

By Eq.~(\ref{eq: gen_green}),
\begin{widetext}
\begin{align} \label{eq: gen_green1}
    G_{1, L}(\omega) & = \frac{A}{B}
    = \frac{(-1)^{L-1} (\beta_{M+2} \cdots \beta_{2M})^{M+1} \prod_{\substack{j \in \{M+2,\ldots,2M\} \\ k \in \{1,\ldots,M+1\} }} (\beta_j -\beta_k)^{-1} [(-1)^{M-1} (-t_M) \beta_{M+2}  \cdots \beta_{2M}]^{L-1}}{(\beta_{M+1} \cdots \beta_{2M})^M \prod_{\substack{j \in \{M+1,\ldots,2M\} \\ k \in \{1,\ldots,M\} }} (\beta_j - \beta_k)^{-1} [(-1)^M (-t_M) \beta_{M+1} \cdots \beta_{2M}]^L} \notag \\
    & = \frac{(-1)^{L-1} \prod_{j \in \{M+2,\ldots,2M\}} (\beta_j - \beta_{M+1})^{-1} (-1)^{(L-1)(M-1)} (-t_M)^{L-1} }{\beta_{M+1}^M \prod_{k \in \{1,\ldots,M\}} (\beta_{M+1} - \beta_k)^{-1} (-1)^{M L} (-t_M)^L \beta_{M+1}^L} \notag \\
    \\
    & = \frac{(-1)^{L-1} \prod_{j \in \{M+2,\ldots,2M\}} (\beta_j - \beta_{M+1})^{-1} (-1)^{L+M+1}}{\beta_{M+1}^M \prod_{k \in \{1,\ldots,M\}} (\beta_{M+1} - \beta_k)^{-1} (-t_M) \beta_{M+1}^L} \notag \\
    & = \frac{(-1)^M \prod_{j \in \{M+2,\ldots,2M\}} (\beta_j - \beta_{M+1})^{-1}}{(-t_M) \prod_{k \in \{1,\ldots,M\}} (\beta_{M+1} - \beta_k)^{-1} \beta_{M+1}^{M+L}}
    = \frac{(-1)^M \prod_{k \in \{1,\ldots,M\}} (\beta_{M+1} - \beta_k)}{(-t_M) \prod_{j \in \{M+2,\ldots,2M\}} (\beta_j - \beta_{M+1})} \beta_{M+1}^{-(M+L)} . \notag
\end{align}
\end{widetext}
We have assumed that $|\beta_{M+1}| < |\beta_{M+2}|$ so that the numerator only has one dominant term. If $|\beta_{M+1}| = |\beta_{M+2}| = \cdots = |\beta_{M+p}| < |\beta_{M+p+1}|$, the $G_{1, L}(\omega)$ is given by 
\begin{widetext}
\begin{align}
    G_{1, L}(\omega) = \sum_{\begin{pmatrix} \beta_{M+1} & \cdots & \beta_{M+p} \\ \beta_{i_1} & \cdots & \beta_{i_p} \end{pmatrix}} \frac{(-1)^M \prod_{k \in \{1,\ldots,M\}} (\beta_{M+1} - \beta_k)}{(-t_M) \prod_{j \in \{M+2,\ldots,2M\}} (\beta_j - \beta_{M+1})} \beta_{M+1}^{-(M+L)} ,
\end{align}
\end{widetext}
where $\beta_{i_1}, \ldots , \beta_{i_p} \in \{ \beta_{M+1} , \ldots , \beta_{M+p} \}$, and the summation is over all possible permutations of $\beta_{M+1} , \ldots , \beta_{M+p}$.

We can compare the analytic formula to the numerical result by applying Eq.~(\ref{eq: gen_green1}) to the case that $M=2$,
\begin{align} \label{eq: exMeq2}
    | G_{1, L}(\omega) | = \frac{|\beta_3-\beta_1| |\beta_3-\beta_2|}{|t_2| |\beta_4-\beta_3|} |\beta_3|^{-(L+2)} .
\end{align}

\begin{figure} 
    \centering
    \includegraphics[width=0.9\columnwidth]{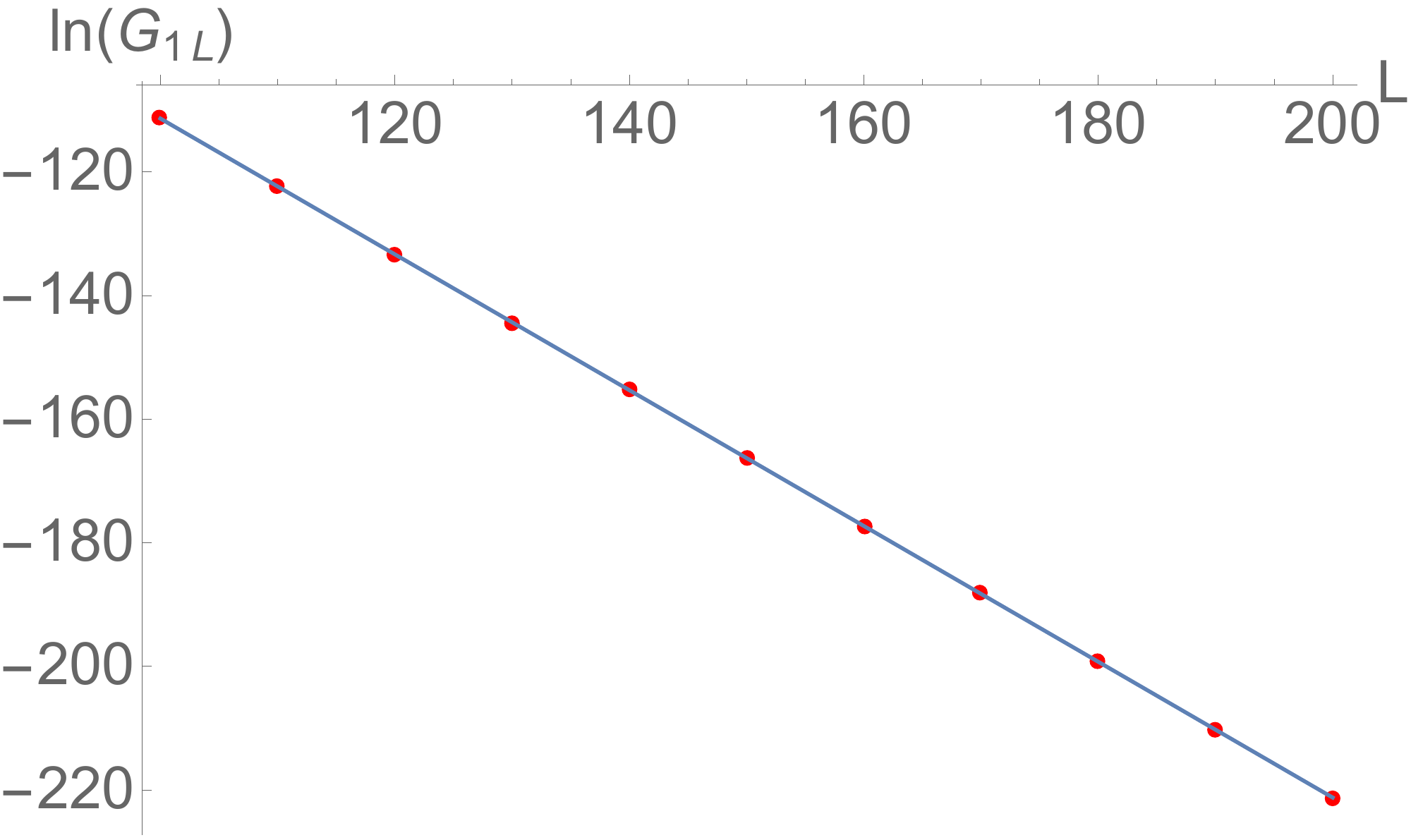}
    \caption{The numerical result and the analytic formula are perfectly aligned. For numerical computation, we choose 10 ``$L$'' points: $L=100,110,120,130,140,150,160,170,180,190,200$. The analytic solution is shown by the straight line.} \label{fig: fit}
\end{figure}

Consider the case that the periodic Hamiltonian $h(\beta)=-(\beta^2-10\beta-50\beta^{-1}+24\beta^{-2})$. Taking $\omega=35$, then $\omega-h(\beta) = \beta^{-2}(\beta ^4-10 \beta ^3+35 \beta ^2-50 \beta +24) = \beta^{-2}(\beta-1)(\beta-2)(\beta-3)(\beta-4)$. Equation (\ref{eq: exMeq2}) gives
\begin{align}
    | G_{1, L}(35) | = 2 \cdot 3^{-(L+2)}.
\end{align}
Figure \ref{fig: fit} shows how closely it matches the numerical result.

Another end-to-end Green's function $G_{L,1}(\omega)$ can be derived in the same way, if $|\beta_1| \leqslant \cdots \leqslant |\beta_{M-1}| < |\beta_{M}| < |\beta_{M+1}| \leqslant \cdots \leqslant |\beta_{2M}|$, it has an asymptotic form
\begin{align}
    & G_{L,1}(\omega) \notag \\
    = & \frac{(-1)^{M} \prod_{j \in \{M+1,\ldots,2M\}} (\beta_{j} - \beta_M) \beta_M^{M+L-2}}{(\beta_{M+1} \cdots \beta_{2M})^2 (-t_M) \prod_{k \in \{1,\ldots,M-1\}} (\beta_M - \beta_{k})}  .
\end{align}
The detailed derivation is in Appendix \ref{sec: another}.

\section{Green's function in the bulk region} \label{sec: main_green_bulk}
\begin{figure*} 
    \centering 
    \subfigure[]{
        \centering
        \includegraphics[width=0.6\columnwidth]{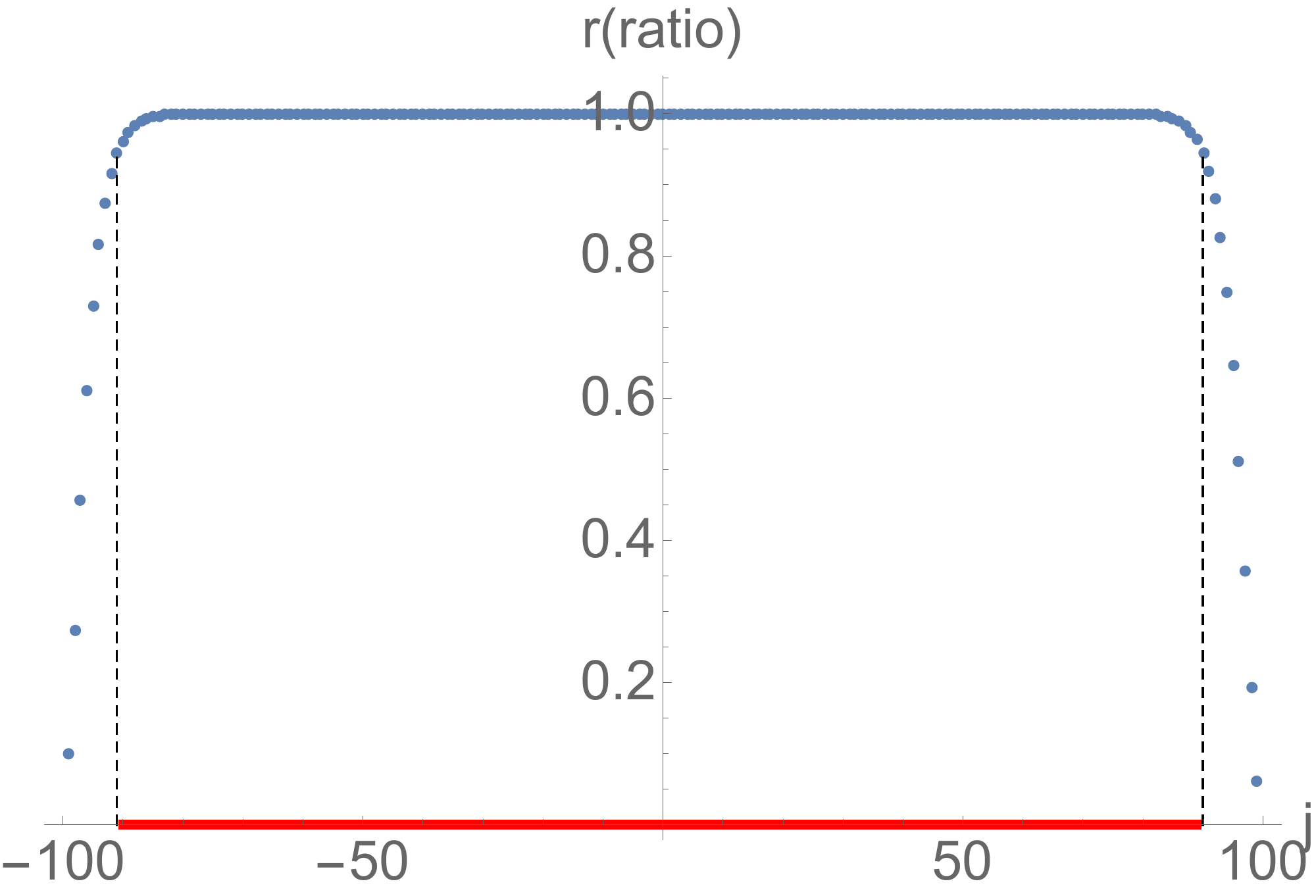}
    }
    \subfigure[]{ 
        \centering
        \includegraphics[width=0.6\columnwidth]{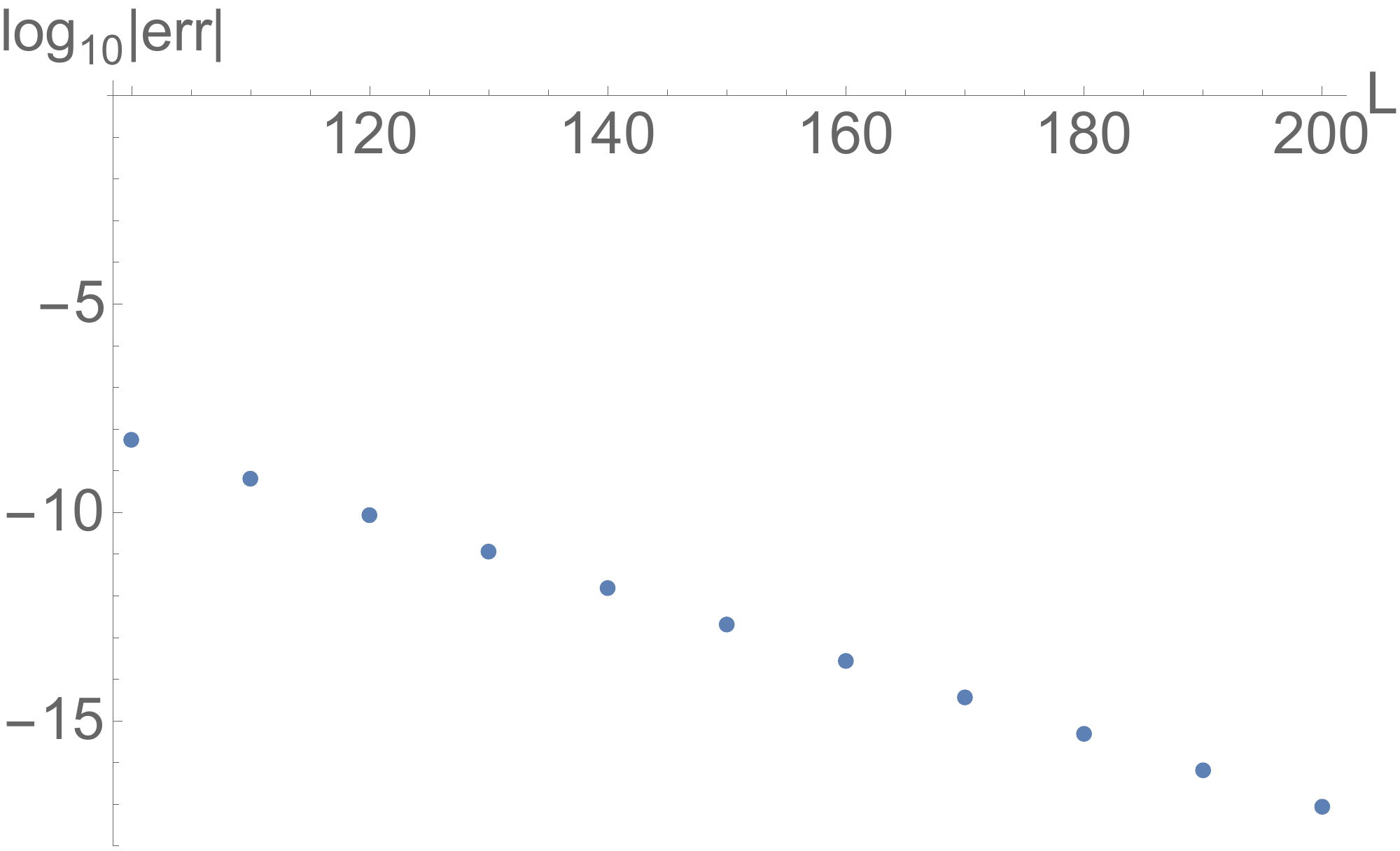}
    } 
    \subfigure[]{
        \centering
        \includegraphics[width=0.6\columnwidth]{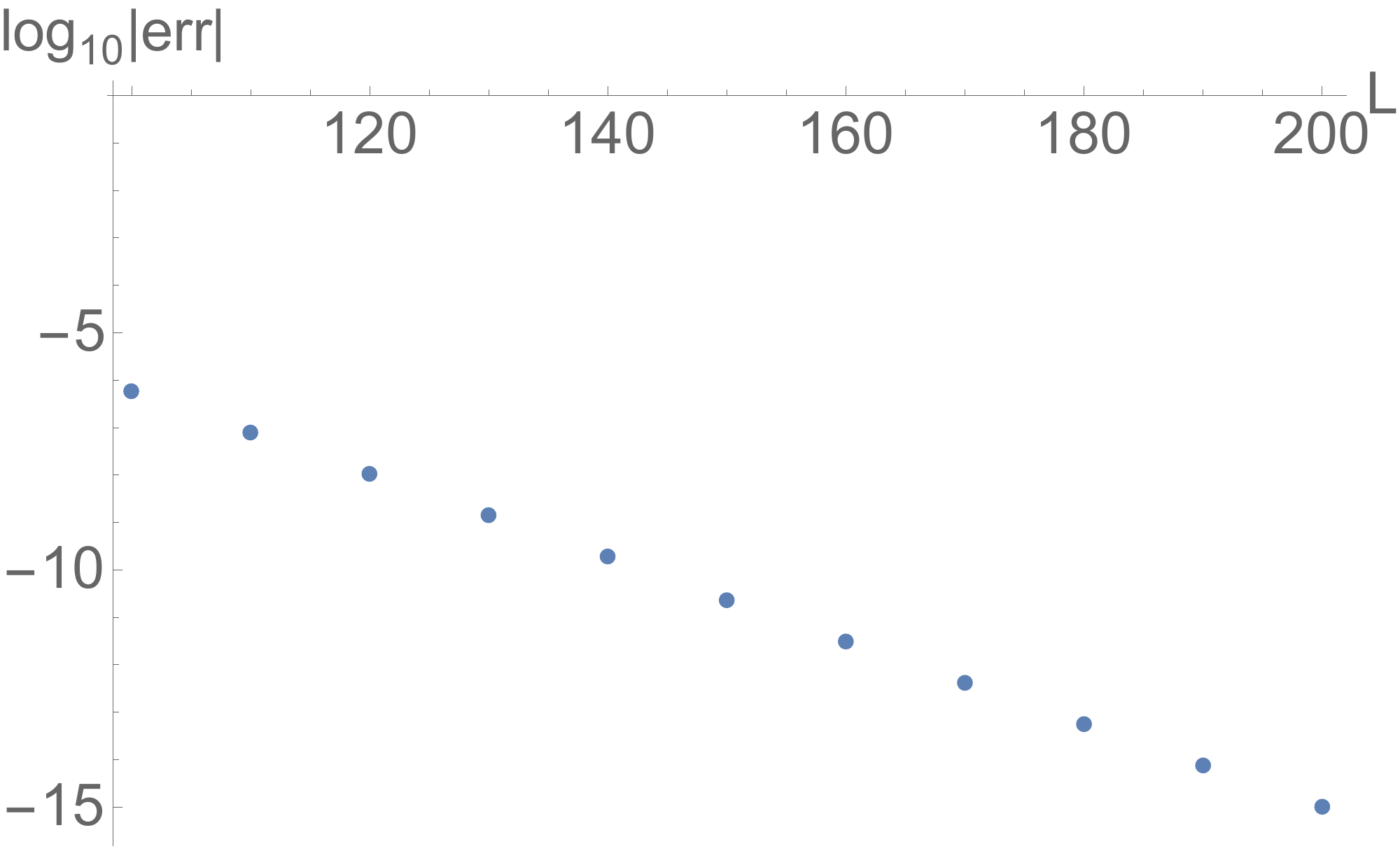}
    } 
    \caption[]{The following three cases has the Hamiltonian $h(\beta)=-(\beta^2-10\beta-50\beta^{-1}+24\beta^{-2})$, and we take $\omega=35$. (a) The ratio $r=\frac{G_{\frac{L}{2}+j,\frac{L}{2}-j}(\omega)}{\int_{\text{GBZ}} \frac{d \beta}{2 \pi i \beta} \frac{\beta^{2j}}{\omega - h(\beta)}}$ against $j$. $G_{\frac{L}{2}+j,\frac{L}{2}-j}(\omega)$ is computed numerically, with the lattice length $L$ set to be $200$. The bulk region is shown as the red line, here we choose the bulk region by choosing sites $j$ at which the ratio is over $0.95$. (b) The error of $G_{\frac{L}{2},\frac{L}{2}}$ as a function of the size of system $L$. (c) The error of $G_{\frac{L}{2}+5,\frac{L}{2}}$ as a function of the size of system $L$. }  \label{fig: error}
\end{figure*}

The prior GBZ-based integral formula can give the exact answer for Green's functions in the bulk region. In mathematical terms, it means for finite $i$ and $j$ value,
\begin{align} \label{eq: green_bulk}
    \lim_{L \rightarrow \infty} G_{\frac{L}{2}+i,\frac{L}{2}+j}(\omega) = \int_{\text{GBZ}} \frac{d \beta}{2 \pi i \beta} \frac{\beta^{i-j}}{\omega - h(\beta)} .
\end{align}
In other words, for large $L$, $G_{\frac{L}{2}+i,\frac{L}{2}+j}(\omega)$ can be expressed as the above GBZ-based integral formula if $i$ and $j$ are smaller than $\frac{L}{2}$, i.e., two sites $L/2+i$ and $L/2+j$ are away from the boundary. Note that when the end-to-end Green's function, say $G_{1,L}(\omega)$, is considered, this criteria is not met, which can be illustrated by plotting the ratio $r = \frac{G_{\frac{L}{2}+j,\frac{L}{2}-j}(\omega)}{\int_{\text{GBZ}} \frac{d \beta}{2 \pi i \beta} \frac{\beta^{2j}}{\omega - h(\beta)}}$ against $j$ as shown in Fig.~\ref{fig: error}(a). The ratio $r$ is approximately unity when the two sites $L/2+j$ and $L/2-j$ are away from the boundary, and we call this region the bulk region, which is shown as the red line in Fig.~\ref{fig: error}(a). Away from the bulk region, the Green's functions largely deviate from the result given by the GBZ-based integral formula.

In Appendix \ref{sec: proof_bulk}, we prove Eq.~(\ref{eq: green_bulk}) by using an estimation constructed in Appendix \ref{sec: estimate} using the Widom's formula. Consider the case that the periodic Hamiltonian $h(\beta)=-(\beta^2-10\beta-50\beta^{-1}+24\beta^{-2})$ and $\omega=35$. The error of a Green's function is defined as the difference between the Green's function of a finite system size and the RHS of Eq.~(\ref{eq: green_bulk}), i.e. $\text{error}(L) := G_{\frac{L}{2}+i,\frac{L}{2}+j}(\omega) - \int_{\text{GBZ}} \frac{d \beta}{2 \pi i \beta} \frac{\beta^{i-j}}{\omega - h(\beta)}$. Figures \ref{fig: error}(b) and \ref{fig: error}(c) display the error of Green's function as a function of system size $L$, demonstrating that the error decreases exponentially as system size increases. This exponentially decreasing behavior of the error is not a coincidence, in Appendix \ref{sec: proof_bulk}, we show that the speed at which the Green's functions in the bulk region approach the prior established integral formula is not slower than an exponential decay as the system size increases and explains why it fails near the boundary.

\section{Signal amplification and NHSE} \label{sec: NHSE}
In this section, we discuss the usage of the end-to-end Green's functions on the detection of the NHSE. We conclude that the signal amplification implies the occurrence of the NHSE in single-band systems. This is similar to the result of Refs.~\cite{PhysRevLett.126.216407, PhysRevB.103.195157} that exponentially growing bulk Green's function implies the occurrence of the NHSE in single-band systems. In addition, our result shows that this result not only holds for the bulk Green's functions (the bulk Green's functions in Ref.~\cite{PhysRevB.103.195157} is corresponding to the Green's functions in the bulk region as discussed in the previous section) but also holds for the end-to-end Green's functions.

In the large $L$ limit, the end-to-end Green's functions have asymptotic formulas 
\begin{align}
    G_{1,L}(\omega) & = a_1 [\beta_M(\omega)]^L, \notag \\
    G_{L,1}(\omega) & = a_2 [\beta_{M+1}(\omega)]^{-L} ,
\end{align}
where $\beta_{j=1,\ldots,2M}(\omega)$ are roots of $\omega-h(\beta) = 0$ ordered as $|\beta_1(\omega)|\leqslant \cdots \leqslant |\beta_{2M}(\omega)|$, and $a_1$, $a_2$ are two coefficients that can be expressed in terms of $\beta_{j=1,\ldots,2M}(\omega)$ and hopping parameters.

Let us first recall the relation between the spectral winding number and the NHSE. The spectral winding number of $h(\beta)$ with respect to the reference energy $\omega$ is defined by 
\begin{align} \label{eq: winding}
    w_{\omega}[h(\beta)] & = \int_{|\beta|=1} \frac{1}{2 \pi i} d\ln[h(\beta)-\omega] \notag \\
    & = - M + \sum_{|\beta_j| < 1 } 1 ,
\end{align}
where $\beta_{j=1,\ldots,2M}(\omega)$ are roots of $\omega-h(\beta) = 0$ ordered as above. Eq.~(\ref{eq: winding}) implies that the spectral winding number counts the number of the roots of $\omega-h(\beta) = 0$ encircled by the unit circle minus $M$. The NHSE is presented in a system which has an intrinsic non-Hermitian point-gap topology \cite{origin2020,PhysRevResearch.1.023013}. In other words, if $w_{\omega}[h(\beta)] \neq 0$ for some reference energy $\omega$, the system has the NHSE \cite{origin2020}.

If the input signal with a frequency $\omega$ at the last site is amplified at the first site, which means that $G_{1,L}(\omega) \gg 1$ and $|\beta_M(\omega)|>1$, then the number of the roots of $\omega-h(\beta) = 0$ encircled by the unit circle is less than $M$, and by Eq.~(\ref{eq: winding}) the spectral winding number of $h(\beta)$ with respect to the reference energy $\omega$ is nonzero. If the input signal with a frequency $\omega$ at the first site is amplified at the last site, which means that $G_{L,1}(\omega) \gg  1$ and $|\beta_{M+1}(\omega)|<1$, then the number of the roots of $\omega-h(\beta) = 0$ encircled by the unit circle is more than $M$, and by Eq.~(\ref{eq: winding}) the spectral winding number of $h(\beta)$ with respect to the reference energy $\omega$ is nonzero. Conversely, if the spectral winding number of $h(\beta)$ with respect to the reference energy $\omega$ is nonzero, then either $|\beta_M(\omega)|>1$ or $|\beta_{M+1}(\omega)|<1$, implying $G_{1,L}(\omega) \gg 1$ or $G_{L,1}(\omega) \gg  1$. Thus the spectral winding number of $h(\beta)$ with respect to the reference energy $\omega$ is non-zero if and only if $G_{1,L}(\omega) \gg 1$ or $G_{L,1}(\omega) \gg  1$. The above discussion builds a correspondence between the signal amplification at a frequency $\omega$ and the spectral winding number of $h(\beta)$ with respect to the reference energy $\omega$ in 1D single-band systems. By the relation between the spectral winding number and the NHSE, we conclude that single-band systems with the NHSE must amplify a signal with some frequency $\omega$.

The above relation between the end-to-end Green's functions and the NHSE gives a potential approach to detect the NHSE in experiments, which is to measure the signal amplification from one end to another end for different complex frequencies (the real part of the complex frequency represents the resonant frequency and the imaginary part of the complex frequency represents a finite linewidths at the resonant frequency).

\section{Conclusions} \label{sec: con}
In this paper, we obtain the exact formula for the end-to-end Green's function and the accurate asymptotic formula. We believe that these exact formulas make it possible to directly and accurately compare theoretical results with quantitative experimental measurable quantities. In practice, our results will aid future directional amplification tests and directional amplifier design. Furthermore, we verify that the GBZ-based integral Green's function formula in the bulk region to agree with the result in our framework. Our method of calculating the determinant of Toeplitz matrices is based on the Widom's formula, and this approach naturally gives a demonstration of the GBZ-based integral formula for calculating Green's function in the bulk region. Furthermore, we find that the speed at which the Green's functions in the bulk region approach the prior established integral formula is not slower than an exponential decay as the system size increases. This study strengthens the idea that there is a correspondence between the the Green's function and the NHSE. 

\begin{acknowledgments}
    The authors thank Yongxu Fu, Shuxuan Wang, Zhiwei Yin and Jihan Hu for discussions. This work was supported by NSFC Grant No.11275180.
\end{acknowledgments}

\appendix

\section{Tight-binding model including next nearest neighborhood hoppings} \label{sec: range_two}
In this Appendix, we give the formula of the $(1, L)$ entry of the Green's function of the tight-binding model whose hopping range $M=2$. Unlike the asymptotic formula obtained in the main text, this formula is exact for any finite size $L$, even when $L$ is small.

Using Eq.~(\ref{eq: det_toe}) and Eq.~(\ref{eq: det_toe_detail}), the $(1, L)$ entry of the Green's function is given by
\begin{align}
    G_{1, L}(\omega) & = \frac{A}{B} ,
\end{align}
where
\begin{align}
    A = & (-1)^{L-1} \notag \\
    \{ & \beta_4^3 (\beta_4 - \beta_1)^{-1} (\beta_4 - \beta_2)^{-1} (\beta_4 - \beta_3)^{-1} (t_2 \beta_4)^{L-1} \notag \\
    & + \beta_3^3 (\beta_3 - \beta_1)^{-1} (\beta_3 - \beta_2)^{-1} (\beta_3 - \beta_4)^{-1} (t_2 \beta_3)^{L-1} \notag \\
    & + \beta_2^3 (\beta_2 - \beta_1)^{-1} (\beta_2 - \beta_3)^{-1} (\beta_2 - \beta_4)^{-1} (t_2 \beta_2)^{L-1} \notag \\
    & + \beta_1^3 (\beta_1 - \beta_2)^{-1} (\beta_1 - \beta_3)^{-1} (\beta_1 - \beta_4)^{-1} (t_2 \beta_1)^{L-1} \} , \notag
\end{align}
and
\begin{align}
    B = & (\beta_3 \beta_4)^2 \prod_{\substack{j \in \{3, 4\} \\ k \in \{1, 2\} }} (\beta_j - \beta_k)^{-1} (-t_2 \beta_3 \beta_4)^L \notag \\
    & + (\beta_2 \beta_4)^2 \prod_{\substack{j \in \{2, 4\} \\ k \in \{1, 3\} }} (\beta_j - \beta_k)^{-1} (-t_2 \beta_2 \beta_4)^L \notag \\
    & + (\beta_1 \beta_4)^2 \prod_{\substack{j \in \{1, 4\} \\ k \in \{2, 3\} }} (\beta_j - \beta_k)^{-1} (-t_2 \beta_1 \beta_4)^L \notag \\
    & + (\beta_2 \beta_3)^2 \prod_{\substack{j \in \{2, 3\} \\ k \in \{1, 4\} }} (\beta_j - \beta_k)^{-1} (-t_2 \beta_2 \beta_3)^L \notag \\
    & + (\beta_1 \beta_3)^2 \prod_{\substack{j \in \{1, 3\} \\ k \in \{2, 4\} }} (\beta_j - \beta_k)^{-1} (-t_2 \beta_1 \beta_3)^L \notag \\
    & + (\beta_1 \beta_2)^2 \prod_{\substack{j \in \{1, 2\} \\ k \in \{3, 4\} }} (\beta_j - \beta_k)^{-1} (-t_2 \beta_1 \beta_2)^L . \notag
\end{align}
We can see from the above equation that for a system with a finite size $L$, all zeros of the function $t_2 \beta^2 + t_1 \beta + (t_0 - \omega) + t_{-1} \beta^{-1} + t_{-2} \beta^{-2}$ contribute to the $(1, L)$-entry of the Green's function.

\section{Derivation of another end-to-end Green's function} \label{sec: another}
In the main text, we obtain the $(1, L)$-entry of the Green's function, while in this Appendix, we derive a formula for another end-to-end Green's function, i.e., the $(L, 1)$ entry of the Green's function.

The $(L, 1)$ entry of the Green's function is
\begin{align} \label{eq: gen_green_1}
    & G_{L, 1}(\omega) \notag \\
    = & (\omega-H)^{-1}_{L, 1} \notag \\
    = & \frac{(-1)^{L-1} \underbrace{\begin{vmatrix} 
        -t_{-1} & \omega-t_0 & \cdots & -t_M & \cdots & 0 \\
        -t_{-2} & -t_{-1} & \omega-t_0 & \cdots & \cdots & 0 \\
        \vdots & \ddots & \ddots & \ddots & \ddots & \vdots \\
        -t_{-M} & \cdots & \cdots & \cdots & \cdots & \cdots \\
        \vdots & \ddots & \ddots & \ddots & \ddots & \vdots \\
        0 & \cdots & 0 & -t_{-M} & \cdots & -t_{-1}
    \end{vmatrix}}_{L-1}}{\det(\omega-H)} .
\end{align}

Applying the Widom's formula, if $|\beta_1| \leqslant \cdots \leqslant |\beta_{M-1}| < |\beta_{M}| < |\beta_{M+1}| \leqslant \cdots \leqslant |\beta_{2M}|$, the dominant term of the numerator is given by 
\begin{align}
    A = (-1)^{L-1} c_A \omega_A^{L-1} ,
\end{align}
where 
\begin{align}
    \omega_A & = (-1)^{M+1}(-t_M) \beta_{M} \beta_{M+1} \cdots \beta_{2M} , \notag \\
    c_A & = (\beta_{M} \beta_{M+1} \cdots \beta_{2M})^{M-1} \prod_{\substack{j \in \{M,\ldots,2M\} \\ k \in \{1,\ldots,M-1\} }} (\beta_j - \beta_k)^{-1}.
\end{align}
The dominant term of the denominator is the same as Eq.~(\ref{eq: denom}).
Hence
\begin{widetext}
    \begin{align} \label{eq: gl1}
        G_{L, 1}(\omega) & = \frac{A}{B} \notag \\
        & = \frac{(-1)^{L-1} (\beta_{M} \beta_{M+1} \cdots \beta_{2M})^{M-1} \prod_{\substack{j \in \{M,\ldots,2M\} \\ k \in \{1,\ldots,M-1\} }} (\beta_j - \beta_k)^{-1} [(-1)^{M+1}(-t_M) \beta_{M} \beta_{M+1} \cdots \beta_{2M}]^{L-1}}{(\beta_{M+1} \cdots \beta_{2M})^M \prod_{\substack{j \in \{M+1,\ldots,2M\} \\ k \in \{1,\ldots,M\} }} (\beta_j - \beta_k)^{-1} [(-1)^M (-t_M) \beta_{M+1} \cdots \beta_{2M}]^L} \notag \\
        & = \frac{(-1)^{L-1} \beta_M^{M-1} \prod_{k \in \{1,\ldots,M-1\}} (\beta_M - \beta_{k})^{-1} (-1)^{(L-1)(M+1)} (-t_M)^{L-1} \beta_M^{L-1}}{ \prod_{j \in \{M+1,\ldots,2M\}} (\beta_{j} - \beta_M)^{-1} (-1)^{M L} (-t_M)^L (\beta_{M+1} \cdots \beta_{2M})^2} \notag \\
        & = \frac{(-1)^{M} \beta_M^{M+L-2} \prod_{k \in \{1,\ldots,M-1\}} (\beta_M - \beta_{k})^{-1}}{(\beta_{M+1} \cdots \beta_{2M})^2 (-t_M) \prod_{j \in \{M+1,\ldots,2M\}} (\beta_{j} - \beta_M)^{-1}} \notag \\
        & = \frac{(-1)^{M} \prod_{j \in \{M+1,\ldots,2M\}} (\beta_{j} - \beta_M) }{(\beta_{M+1} \cdots \beta_{2M})^2 (-t_M) \prod_{k \in \{1,\ldots,M-1\}} (\beta_M - \beta_{k})} \beta_M^{M+L-2} .
    \end{align}
\end{widetext}

\begin{figure} 
    \centering
    \includegraphics[width=0.9\columnwidth]{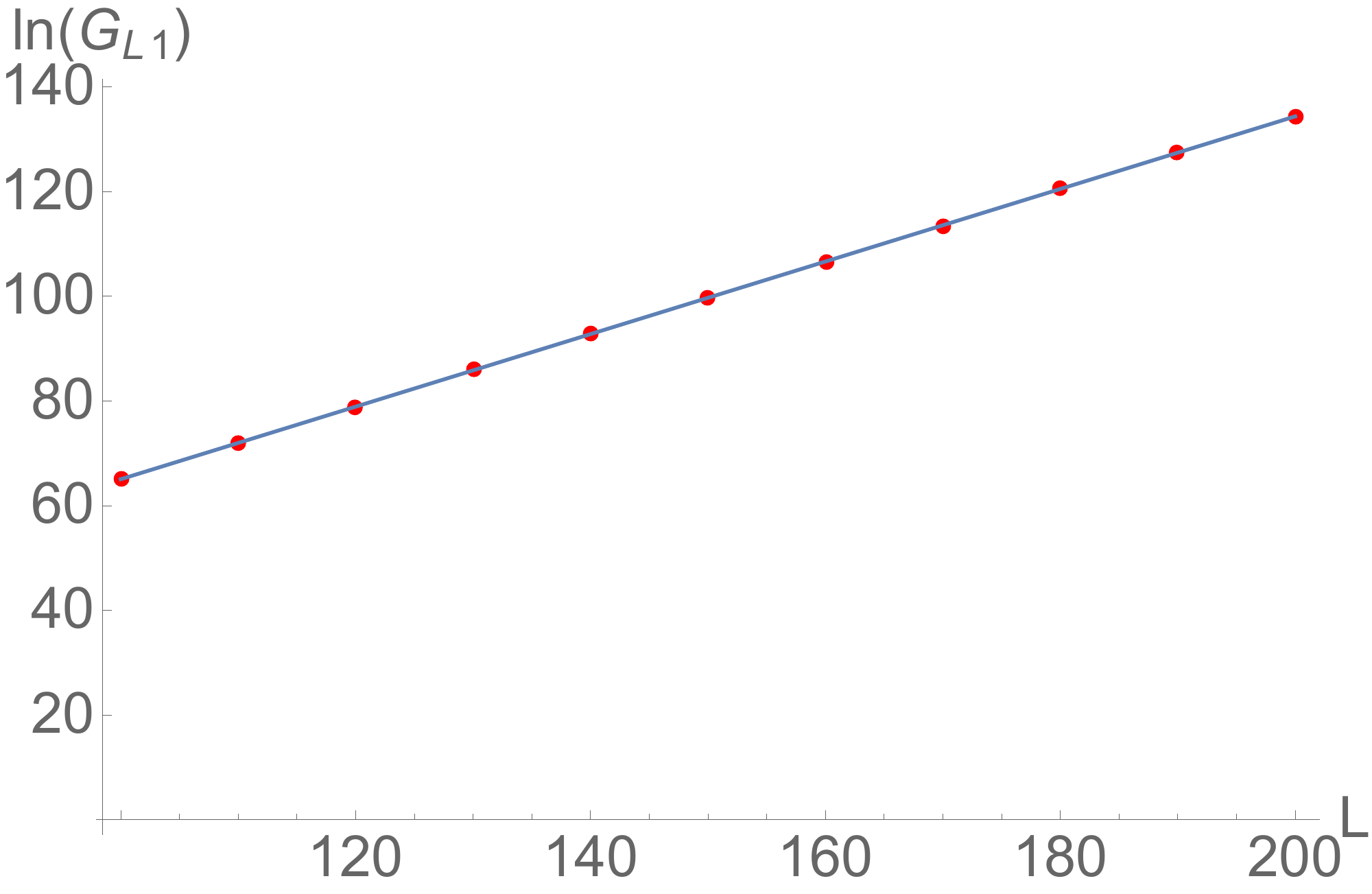}
    \caption{The numerical result and the analytic formula are perfectly aligned. For numerical computation, we choose 10 ``$L$'' points: $L=100,110,120,130,140,150,160,170,180,190,200$. The analytic solution is shown by the straight line.} \label{fig: fit1}
\end{figure}

We test Eq.~(\ref{eq: gl1}) by the case that $h(\beta) = -(\beta^2-10\beta-50\beta^{-1}+24\beta^{-2})$ and $\omega=35$. By Eq.~(\ref{eq: gl1}), it is easy to obtain that $G_{L, 1}(35) = \frac{1}{72} 2^L$. This formula matches with the numerical result as shown in Fig.~\ref{fig: fit1}. 

\section{Diagonal elements of the Green's function} \label{sec: diag}
In this Appendix, we discuss the diagonal elements of Green's function relating to each site's self-energy. We focus on sites $1$ and $L$, which are at the end of the 1D chain. The $(1,1)$- and $(L,L)$- entry of Green's function are given by 
\begin{align} \label{eq: green_diag}
    G_{1,1}(\omega) = G_{L,L}(\omega) = \frac{\Delta_{L-1}}{\Delta_{L}}, 
\end{align}
where $\Delta_{n} = \det[(\omega-H)_{n \times n}]$. This formula can be compute by using Eq.~(\ref{eq: det_toe}) in systems with any finite size $L$. Here we consider the thermodynamics limit of Eq.~(\ref{eq: green_diag}), i.e., we take $L \to \infty$,
\begin{align} \label{eq: green_diag_true}
    \lim_{L \to \infty} G_{1,1}(\omega) & = \frac{c_B [(-1)^M (-t_M) \beta_{M+1} \cdots \beta_{2M}]^{L-1}}{c_B [(-1)^M (-t_M) \beta_{M+1} \cdots \beta_{2M}]^{L}} \notag \\
    & = \frac{1}{(-1)^{M+1} t_M \beta_{M+1} \cdots \beta_{2M}} .
\end{align}

\section{Green's function in the bulk region} \label{sec: proof_bulk}
The GBZ-based integral formula for Green's function in Ref.~\cite{PhysRevB.103.L241408} is accurate in the bulk region, i.e., for finite $i$ and $j$ value,
\begin{align} \label{eq: bulk_formula}
    \lim_{L \rightarrow \infty} G_{\frac{L}{2}+i,\frac{L}{2}+j}(\omega) = \int_{\text{GBZ}} \frac{d \beta}{2 \pi i \beta} \frac{\beta^{i-j}}{\omega - h(\beta)} .
\end{align}
As Ref.~\cite{PhysRevB.103.L241408} points out, the GBZ curve and the circle with radius R encompass the same set of poles of $\frac{1}{2 \pi i \beta} \frac{\beta^{i-j}}{\omega - h(\beta)}$, hence this formula is identical to
\begin{align} \label{eq: bulk_formula1}
    \lim_{L \rightarrow \infty} G_{\frac{L}{2}+i,\frac{L}{2}+j}(\omega) = \int_{|\beta|=R} \frac{d \beta}{2 \pi i \beta} \frac{\beta^{i-j}}{\omega - h(\beta)} ,
\end{align}
where $R$ is chosen to satisfy $|\beta_{M}| < R < |\beta_{M+1}|$, and $\beta_{j=1,\ldots,2M}$ are roots of $\omega-h(\beta) = 0$ ordered as $|\beta_1|\leqslant \cdots \leqslant |\beta_{2M}|$. In the following, we prove Eq.~(\ref{eq: bulk_formula1}).

First, we formulate Eq.~(\ref{eq: bulk_formula1}) in an alternative way. We introduce a matrix $V_L^R$,
\begin{align}
    (V_L^R)_{ij} = \frac{1}{\sqrt{L}}(R \zeta_L^{i-1})^{-(j-1)}, 
\end{align}
where $R$ is the radius of a circle, $L$ is the length of the chain and the size of $V_L^R$, and $\zeta_L = \exp(2 \pi i/L)$ is a primitive $L^{\text{th}}$ root of $1$. Due to
\begin{align}
    & \sum_{k=1}^L \frac{1}{\sqrt{L}} (R \zeta_L^{k-1})^{i-1} \frac{1}{\sqrt{L}} (R \zeta_L^{k-1})^{-(j-1)} \notag \\
    = & \frac{1}{L} \sum_{k=1}^L \zeta_L^{(k-1)(i-j)} R^{i-j} \notag \\
    = & \frac{1}{L} L \delta_{i,j} R^{i-j} \\
    = & \delta_{i,j} \notag,
\end{align}
$(V_L^R)^{-1}_{i,j} = \frac{1}{\sqrt{L}} (R \zeta_L^{j-1})^{i-1}$. We generalize the definition of the circulant matrix \cite{bottcher2005spectral, maa/1119019654} to define the circulant matrix with radius $R$ and size $L$ by 
\begin{align}
    C_L^R(a) = (V_L^R)^{-1} \text{diag}(a(R),a(R \zeta_L), \ldots, a(R \zeta_L^{L-1})) V_L^R,
\end{align}
where $a(\beta)$ is the symbol of the circulant matrix. The inverse matrix $(C_L^R)^{-1}(a)$ is again a circulant matrix:
\begin{align}
    (C_L^R)^{-1}(a) = (V_L^R)^{-1} \text{diag}(\frac{1}{a(R)}, \ldots, \frac{1}{a(R \zeta_L^{L-1})}) V_L^R.
\end{align}
Equation (\ref{eq: bulk_formula1}) is equivalent to the following statement,
\begin{align}
    \lim_{L \rightarrow \infty} [T_L^{-1}(a) - (C_L^R)^{-1}(a)]_{\frac{L}{2}+i,\frac{L}{2}+j} = 0,
\end{align}
where $i$ and $j$ are two fixed integers, and the radius $R$ is chosen such that half number of zeros of $a(\beta)=\omega-h(\beta)$ are enclosed by the circle of radius $R$.

The symbol $a(\beta)$ of the Toeplitz matrices and the circulant matrices considered here is taken to be Laurent polynomial 
\begin{align}
    a(\beta) = \sum_{j=-r}^s a_j \beta^j .
\end{align}
The Toeplitz matrix with symbol $a(\beta)$ is defined by 
\begin{align}
    [T_L(a)]_{i,j} = a_{j-i} .
\end{align}
The circulant matrix with symbol $a(\beta)$ and radius $R$ is given by 
\begin{align}
    [C_L^R(a)]_{i,j} & = \sum_{k=1}^L (V_L^R)^{-1}_{i,k} a(R \zeta_L^{k-1}) (V_L^R)_{k,j} \notag \\
    & = \frac{1}{L} \sum_{k=1}^L \zeta_L^{(k-1)(i-j)} R^{i-j} a(R \zeta_L^{k-1}) \notag \\
    & = \frac{1}{L} \sum_m \sum_{k=1}^L \zeta_L^{(k-1)(i-j)} R^{i-j} a_m R^m \zeta_L^{m(k-1)} \notag \\
    & = \sum_m \delta_{(m+i-j \equiv 0 \mod L)} a_m R^{m+i-j} .
\end{align}
For the case that $i-j$ is much smaller than the size $L$ of system, $[C_L^R(a)]_{i,j} = a_{j-i} = [T_L(a)]_{i,j}$, meanwhile, for the case that $i-j$ is comparable to $L$, if $i-j=L-m$, $[C_L^R(a)]_{i,j} = a_{m} R^L$, if $i-j=-L-m$, $[C_L^R(a)]_{i,j} = a_{m} R^{-L}$. Hence, elements near the lower left corner or the upper right corner of the circulant matrix is amplified or suppressed by $R^L$ factor.

In our next discussion, we will require the following estimate, which we will develop in Appendix \ref{sec: estimate}. For any $i$, there exists constants $c_1$ and $c_2$ such that
\begin{align} \label{eq: est}
    |[T_L(a)]^{-1}_{j,\frac{L}{2}+i}| & \leqslant c_1 |\beta_{M+1}(\omega)|^{-\frac{L}{2}}, \notag \\
    |[T_L(a)]^{-1}_{L+1-j,\frac{L}{2}+i}| & \leqslant c_2 |\beta_M(\omega)|^{\frac{L}{2}},
\end{align}
where $\beta_{j=1,\ldots,2M}$ are roots of $a(\beta) = 0$ ordered as $|\beta_1|\leqslant \cdots \leqslant |\beta_{2M}|$, and $1\leqslant j \leqslant \max(r,s)=M$.

Inequality
\begin{widetext}
\begin{align} \label{eq: con_speed}
    & |[T_L^{-1}(a) - (C_L^R)^{-1}(a)]_{\frac{L}{2}+i,\frac{L}{2}+j} |\notag \\
    = & | \{ (C_L^R)^{-1}(a) [C_L^R(a) - T_L(a)] T_L^{-1}(a) \}_{\frac{L}{2}+i,\frac{L}{2}+j} | \notag \\
    = & | [(C_L^R)^{-1}(a)]_{\frac{L}{2}+i,l} [C_L^R(a) - T_L(a)]_{l,L+1-k} [T_L^{-1}(a)]_{L+1-k,\frac{L}{2}+j} \notag \\
    & + (C_L^R)^{-1}(a)]_{\frac{L}{2}+i,L+1-l} [C_L^R(a) - T_L(a)]_{L+1-l,k} [T_L^{-1}(a)]_{k,\frac{L}{2}+j} | \notag \\
    \leqslant & |[(C_L^R)^{-1}(a)]_{\frac{L}{2}+i,l}| \, |[C_L^R(a) - T_L(a)]_{l,L+1-k}| \, |[T_L^{-1}(a)]_{L+1-k,\frac{L}{2}+j}| \notag \\
    & + |(C_L^R)^{-1}(a)]_{\frac{L}{2}+i,L+1-l}| \, |[C_L^R(a) - T_L(a)]_{L+1-l,k}| \, |[T_L^{-1}(a)]_{k,\frac{L}{2}+j}| \notag \\
    \leqslant & b_1 R^{\frac{L}{2}} R^{-L} c_2 |\beta_{M}|^{\frac{L}{2}}  + b_2 R^{-\frac{L}{2}} R^L c_1 |\beta_{M+1}|^{-\frac{L}{2}} \notag \\
    \leqslant & b_1 c_2 (\frac{|\beta_{M}|}{R})^{\frac{L}{2}} + b_2 c_1 (\frac{R}{|\beta_{M+1}|})^{\frac{L}{2}} 
\end{align}
\end{widetext}
implies that 
\begin{align}
    \lim_{L \rightarrow \infty} |[T_L^{-1}(a) - (C_L^R)^{-1}(a)]_{\frac{L}{2}+i,\frac{L}{2}+j} = 0 ,
\end{align}
since $\frac{|\beta_{M}|}{R}, \frac{R}{|\beta_{M+1}|} < 1$. The preceding proof also demonstrates why $R$ should be chosen to satisfy $|\beta_{M}| < R < |\beta_{M+1}|$.

Equation~(\ref{eq: con_speed}) also explains the deviation of the Green's functions near the boundary from the GBZ integral based formula as shown in Fig.~\ref{fig: error}. (a). Before Eq.~(\ref{eq: con_speed}), we have assumed that $i$,$j$ are fixed and the size $L$ tends to infinity, which means that $i$,$j$ are small compared with $L/2$. When the Green's functions near the boundary are considered, $i$,$j$ in $G_{\frac{L}{2}+i,\frac{L}{2}+j}(\omega)$ are comparable to $L/2$ and tend to infinity together with $L$, the last two rows of Eq.~(\ref{eq: con_speed}) should be modified to
\begin{align}
    & |[T_L^{-1}(a) - (C_L^R)^{-1}(a)]_{\frac{L}{2}+i,\frac{L}{2}+j} |\notag \\
    \leqslant &  b_1 R^L R^{-L} c_2 |\beta_{M}|^{\frac{L}{2}}  + b_2 R^{-L} R^L c_1 |\beta_{M+1}|^{-\frac{L}{2}} \notag \\
    = & b_1 c_2 |\beta_{M}|^{\frac{L}{2}} + b_2 c_1 |\beta_{M+1}|^{-\frac{L}{2}},
\end{align}
which no longer converges to zero as $L$ tends to infinity in general.

Recall that Figs.~\ref{fig: error}(b) and \ref{fig: error}(c) display the difference between the Green's function $G_{\frac{L}{2}+i,\frac{L}{2}+j}(\omega)$ in the bulk region and the integral formula $\int_{\text{GBZ}} \frac{d \beta}{2 \pi i \beta} \frac{\beta^{i-j}}{\omega - h(\beta)}$ as a function of system size $L$, demonstrating that the difference decreases exponentially as system size increases. It can be explained by Eq.~(\ref{eq: con_speed}) and we conclude that the speed at which the Green's functions in the bulk region approach the integral formula is not slower than an exponential decay as the system size increases.

\section{Asymptotic estimate} \label{sec: estimate}
We build an estimate (\ref{eq: est}) in this Appendix, i.e., for any $i$, there exists constants $c_1$ and $c_2$ such that
\begin{align}
    |[T_L(a)]^{-1}_{j,\frac{L}{2}+i}| & \leqslant c_1 |\beta_{M+1}(\omega)|^{-\frac{L}{2}}, \notag \\
    |[T_L(a)]^{-1}_{L+1-j,\frac{L}{2}+i}| & \leqslant c_2 |\beta_M(\omega)|^{\frac{L}{2}},
\end{align}
where $\beta_{j=1,\ldots,2M}$ are roots of $a(\beta) = 0$ ordered as $|\beta_1|\leqslant \cdots \leqslant |\beta_{2M}|$, and $1\leqslant j \leqslant \max(r,s)=M$.

Consider the banded Toeplitz matrix with symbol $a(\beta) = \sum_{j=-M}^M a_j \beta^j$,
\begin{align}
    T_L(a) = \begin{pmatrix}
        a_0 & a_1 & \cdots & a_M & 0 & \cdots & 0 \\
        a_{-1} & a_0 & a_1 & \cdots & a_M & 0 & \cdots \\
        \vdots & \ddots & \ddots & \ddots & \ddots & \ddots & \vdots \\
        a_{-M} & a_{-M+1} & \cdots & \cdots & \cdots & \cdots & \cdots \\
        \vdots & \ddots & \ddots & \ddots & \ddots & \ddots & \vdots \\
        0 & \cdots & 0 & a_{-M} & \cdots & a_0 & a_1 \\
        0 & \cdots & 0 & 0 & a_{-M} & \cdots & a_0
    \end{pmatrix}_{L \times L}.
\end{align}
Using Cramer's rule, we obtain
\begin{align} \label{eq: inv_elem}
    [T_L^{-1}(a)]_{1,\frac{L}{2}+k} = (-1)^{\frac{L}{2}+k+1}\frac{\det \begin{pmatrix} A & B \\ C & D \end{pmatrix}}{\det(T_L(a))},
\end{align}
where 
\begin{align}
    A & = \begin{pmatrix}
        a_1 & a_2 & \cdots & a_M & 0 & \cdots & 0 \\
        a_0 & a_1 & a_2 & \cdots & a_{M} & 0 & \cdots \\
        \vdots & \ddots & \ddots & \ddots & \ddots & \ddots & \vdots \\
        a_{-M} & a_{-M+1} & \cdots & \cdots & \cdots & \cdots & \cdots \\
        \vdots & \ddots & \ddots & \ddots & \ddots & \ddots & \vdots \\
        0 & \cdots & 0 & a_{-M} & \cdots & a_0 & a_1
    \end{pmatrix}  , \notag \\
    B & = \begin{pmatrix}
        0 & 0 & \cdots & \cdots & \cdots & \cdots & 0 \\
        \vdots & \ddots & \ddots & \ddots & \ddots & \ddots & \vdots \\
        0 & 0 & \cdots & \cdots & \cdots & \cdots & 0 \\
        a_M & 0 & \cdots & \cdots & \cdots & \cdots & 0 \\
        a_{M-1} & a_M & 0 & \cdots & \cdots & \cdots & 0 \\
        \vdots & \ddots & \ddots & \ddots & \ddots & \ddots & \vdots \\
        a_1 & a_2 & \cdots & a_M & 0 & \cdots & 0
    \end{pmatrix} , \notag \\
    C & = \begin{pmatrix}
        0 & \cdots & 0 & a_{-M} & a_{-M+1} & \cdots & a_{-2} \\
        0 & \cdots & 0 & 0 & a_{-M} & \cdots & a_{-3} \\
        \vdots & \ddots & \ddots & \ddots & \ddots & \ddots & \vdots \\
        0 & \cdots & \cdots & \cdots & \cdots & 0 & a_{-M} \\
        \vdots & \ddots & \ddots & \ddots & \ddots & \ddots & \vdots \\
        0 & \cdots & \cdots & \cdots & \cdots & \cdots & 0 \\
    \end{pmatrix} , \notag \\
    D & = \begin{pmatrix}
        a_0 & a_1 & \cdots & a_M & 0 & \cdots & 0 \\
        a_{-1} & a_0 & a_1 & \cdots & a_M & 0 & \cdots \\
        \vdots & \ddots & \ddots & \ddots & \ddots & \ddots & \vdots \\
        a_{-M} & a_{-M+1} & \ddots & \ddots & \ddots & \ddots & \vdots \\
        \vdots & \ddots & \ddots & \ddots & \ddots & \ddots & \vdots \\
        0 & \cdots & 0 & a_{-M} & \cdots & a_0 & a_1 \\
        0 & \cdots & 0 & 0 & a_{-M} & \cdots & a_0
    \end{pmatrix} ,
\end{align}
and $A$ is a $(\frac{L}{2}+k-1) \times (\frac{L}{2}+k-1)$ matrix, $B$ is a $(\frac{L}{2}+k-1) \times (\frac{L}{2}-k)$ matrix, $C$ is a $(\frac{L}{2}-k) \times (\frac{L}{2}+k-1)$ matrix and $D$ is a $(\frac{L}{2}-k) \times (\frac{L}{2}-k)$ matrix.

The numerator of Eq.~(\ref{eq: inv_elem}) is given by 
\begin{align}
    \det \begin{pmatrix} A & B \\ C & D \end{pmatrix} = \det(A) \det(D - C A^{-1} B) .
\end{align}
Note that in $C A^{-1} B$, only the upper left $(M+1) \times (M+1)$-block is non-zero. Hence, by subtracting $C A^{-1} B$, only the components in the upper left $(M+1) \times (M+1)$-block of $D$ is altered. For convenience, we denote them by $x_1, x_2, \ldots, x_{(M+1)^2}$, and the corresponding components in $D-C A^{-1} B$ by $x'_1, x'_2, \ldots, x'_{(M+1)^2}$, then $\det(D)$ and $\det(D-C A^{-1} B)$ are given by 
\begin{align} \label{eq: compare}
    \det(D) & = \sum_{\alpha} c_{\alpha_1 \cdots \alpha_p}(L) x_{\alpha_1} \cdots x_{\alpha_p}, \notag \\
    \det(D-C A^{-1} B) & = \sum_{\alpha} c_{\alpha_1 \cdots \alpha_p}(L) x'_{\alpha_1} \cdots x'_{\alpha_p} .
\end{align}

According to Widom's formula, $|\det(D)|$ is bounded by $\mathcal{N}_D |\beta_{M+1} \cdots \beta_{2M}|^{\frac{L}{2}}$, where $\mathcal{N}_D$ is a constant independent of $L$. Then, by Eq.~(\ref{eq: compare}), there exists a constant $\mathcal{M}$ that is independent of $L$ such that $|\det(D-C A^{-1} B)| \leqslant \mathcal{M} \mathcal{N}_D |\beta_{M+1} \cdots \beta_{2M}|^{\frac{L}{2}}$. Using the Widom's formula once more, we obtain $|\det(A)| \leqslant \mathcal{N}_A |\beta_{M+2} \cdots \beta_{2M}|^{\frac{L}{2}}$ for a positive $\mathcal{N}_A$ and $|\det(T_L(a))| \geqslant  \mathcal{N} |\beta_{M+1} \cdots \beta_{2M}|^L$ for a positive number $\mathcal{N}$. Inserting these estimates in Eq.~(\ref{eq: inv_elem}) we obtain 
\begin{align}
    |[T_L^{-1}(a)]_{1,\frac{L}{2}+k}| \leqslant \frac{\mathcal{N}_A \mathcal{N}_D \mathcal{M}}{\mathcal{N}} |\beta_{M+1}|^{-\frac{L}{2}}. 
\end{align}

To estimate $|T_L^{-1}(a)]_{i,\frac{L}{2}+k}$ for $i \neq 1$, again we can write
\begin{align} \label{eq: inv_elem1}
    [T_L^{-1}(a)]_{i,\frac{L}{2}+k} = (-1)^{\frac{L}{2}+k+i}\frac{\det \begin{pmatrix} A & B \\ C & D \end{pmatrix}}{\det(T_L(a))}.
\end{align}
In contrast to the case where $i=1$, $A$-block in Eq.~(\ref{eq: inv_elem1}) is not Toeplitz. Let $A = \begin{pmatrix} A_{11} & A_{12} \\ A_{21} & A_{22} \end{pmatrix}$, where $A_{11}$ and $A_{22}$ are Toeplitz. Then $\det(A) = \det(A_{11}) \det(A_{22} - A_{21} A_{11}^{-1} A_{12})$, and the same procedure that was used to estimate $\det \begin{pmatrix} A & B \\ C & D \end{pmatrix}$ may be used to estimate $\det(A)$. Finally, we have
\begin{align}
    |[T_L^{-1}(a)]_{i,\frac{L}{2}+k}| \leqslant \mathcal{N} |\beta_{M+1}|^{-\frac{L}{2}}
\end{align}
for a positive number $\mathcal{N}$.

Similarly, we obtain that
\begin{align}
    |[T_L(a)]^{-1}_{L+1-j,\frac{L}{2}+i}| & \leqslant \mathcal{N} \frac{|\beta_{M} \cdots \beta_{2M}|^{\frac{L}{2}} |\beta_{M+1} \cdots \beta_{2M}|^{\frac{L}{2}}}{|\beta_{M+1} \cdots \beta_{2M}|^L} \notag \\
    & \leqslant \mathcal{N} |\beta_{M}|^{\frac{L}{2}}.
\end{align}

\bibliography{Ref}

\end{document}